\documentclass[twocolumn,showpacs,preprintnumbers,amsmath,amssymb]{revtex4}
\usepackage{graphicx}
\usepackage{dcolumn}
\usepackage{bm}

\begin{document}
\title{Effects of current on vortex and transverse domain walls}
\author{J. He, Z. Li and S. Zhang}
\affiliation{Department of Physics and Astronomy, University of
Missouri-Columbia, Columbia, MO 65211}
\date{\today}

\begin{abstract}
By using the spin torque model in ferromagnets, we compare the
response of vortex and transverse walls to the electrical current.
For a defect-free sample and a small applied current, the steady
state wall mobility is independent of the wall structure. In the
presence of defects, the minimum current required to overcome the
wall pinning potential is much smaller for the vortex wall than for
the transverse wall. During the wall motion, the vortex wall tends
to transform to the transverse wall. We construct a phase diagram
for the wall mobility and the wall transformation driven by the
current.
\end{abstract}

\maketitle

\section{Introduction}

In a magnetic wire, there are two standard walls: transverse wall
(TW) and vortex wall (VW). In a certain range of the wire width and
thickness, both types of the walls are stable
\cite{Miltat1,McMichael}. Experimentally, one can create either type
of the wall by applying a magnetic field in various directions
\cite{Shinjo,Klaui}. The dynamics of both walls driven by a magnetic
field has been extensively simulated \cite{Miltat2, Porter}. In
general, the mobility of the vortex wall is smaller than the
transverse wall, and at a large magnetic field, the vortex core
tends to move toward the wire edge and may be annihilated at the
sample boundary. Thus a transformation from vortex wall to
transverse wall may occur. In this paper, we address the wall
dynamics driven by a current. It would be interesting to see how the
current-driven wall dynamics qualitatively differs from the
field-driven wall dynamics. The present work is also motivated by
recent experimental observations on the domain wall motion
\cite{Klaui}. Particularly, Kl\"{a}ui \emph{et al}. have
quantitatively measured the correlation between wall mobilities and
wall structures, and a transformation from vortex wall to transverse
wall has been clearly identified.

\section{Model}
We start with the generalized Landau-Lifshitz-Gilbert equation which
includes the recently derived spin transfer torques \cite{Zhang}:

\begin{eqnarray}
\label{LLG} \frac{\partial {\bf M}}{\partial t}&=&-\gamma {\bf
M}\times {\bf H}_{eff}+\frac{\alpha}{M_s} {{\bf M}} \times
\frac{\partial {\bf M}}{\partial t}\nonumber\\
&&-\frac{b_{J}}{M_s} {{\bf M}}\times\left({{\bf M}}\times \frac{\partial{\bf
M}}{\partial
x}\right)-\frac{c_{J}}{\emph{M}_{s}}\textbf{M}\times\frac{\partial
\textbf{M}}{\partial x}
\end{eqnarray}

where $\gamma$ is the gyromagnetic ratio, and ${\bf H}_{eff}$ is the
effective magnetic field including the external field, the
anisotropy field, the magnetostatic field, and the exchange field,
$\alpha$ is the Gilbert damping parameter,
$b_{J}=Pj_{e}\mu_{B}/eM_{s}$, $c_{J}=\xi b_{J}$, $\emph{P}$ is the
spin polarization of the current, $j_{e}$ is the current density
along the length direction of nanowire, $\mu_{B}$ is Bohr magneton,
and $\xi$ is a dimensionless constant which describes the degree of
the nonadiabaticity between the spin of the nonequilibrium
conduction electrons and local magnetization.

\section{Calculation procedure}

We have solved the above LLG equation by performing a routine
micromagnetic simulation for a magnetic wire whose geometrical size
is $2{\mu}m$ long (x-direction, also the current direction), $128
nm$ wide, and $8 nm$ thick. The grid size is chosen as
$4\times4\times8 nm^3$. The magnetization at both ends is set to be
along the x-direction and direct inward to the wire; and we use free
boundary conditions for other dimensions. Since the domain wall may
move up to several micrometers in some cases to be discussed below,
it is important to keep the domain wall far away from the ends. To
reduce this end effects due to the finite length of the wire, we
shift the domain wall to the center of the wire after every small
displacement or time interval, corresponding to the geometrical
structure of wire. Using the similar initial domain configurations
as used in \cite{McMichael}, these two types of walls will form and
they are our initial wall configurations at $t=0$. Starting at
$t=0$, a static spin torque or a static magnetic field is applied to
the wall until the end of the simulation. The strength of the spin
torque or the field is chosen in a certain range and the dynamics of
the wall after $t>0$ is reported in the paper. In simulating the
pinning potential, we choose a geometrical boundary roughness by
periodically removing, for every 32 grids in the length direction,
one grid from each edge in the direction of wire width\cite{Jason}.

The material parameters are: the exchange constant $A=
1.3\times10^{-6} erg/cm$, the anisotropy field $H_{K} = 0 (Oe)$, the
saturation magnetization $M_s= 800 emu/cc$, the damping parameter
$\alpha$ is 0.02, the spin polarization $P=0.5$, and the
non-adiabaticity factor $\xi =0.04$.

\section{Results}
\subsection{Defect-free domain wall motion}

A simplest case is a defect-free wire. In this case, the steady
state wall velocity $v_x$ can be analytically derived,
\begin{equation}
v_x = \frac{\gamma H_{e} W}{\alpha} -\frac{c_J}{\alpha}
\end{equation}
where $H_{e}$ it the external field and $W$ is the domain wall width
defined as
\begin{equation}
W^{-1} = \frac{1}{2SM_s^2} \int dV \left( \frac{\partial \bf
M}{\partial x} \right)^2
\end{equation}
where $S$ is the cross section area of the wire. The assumption used
in deriving Eq.~(2) is that the domain wall moves uniformly, i.e.,
${\bf M} = {\bf M}({\bf r}-{v_x}t\hat{\bf x})$, see paper \cite{Li}
of this proceedings. Thus, in the absence of the field, the steady
velocity is $-c_J/\alpha$, independent of the wall
structure \cite{Miltat3}. In the presence of the field, however, the
velocity depends on the wall width defined in Eq.~(3). The vortex
wall has a smaller effective width compared to the transverse wall
and thus the vortex wall has lower velocity \cite{He}. In Fig.~(1),
we show the steady state velocity for the vortex wall and the
transverse wall. The wall velocity is well fitted by Eq.~(2).

\begin{figure}
\centering
\includegraphics[width=8.5cm]{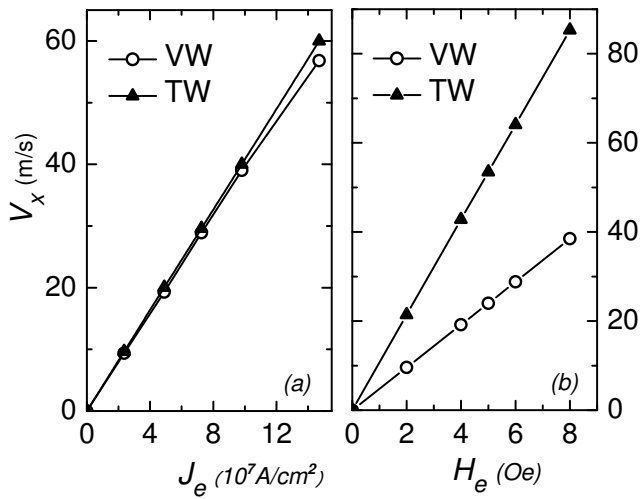}
\caption{ Domain-wall steady state velocity $\it{v_{x}}$ as a
function of (a) current $\it{J_e}$, and (b) field $\it{H_{e}}$. In
(a) $\it{v_{x}}=-c_J/\alpha$ and in (b)
$\it{v_{x}}=\gamma\it{H_{e}}W/\alpha$, both agree with the
analytical result in (2)}
\end{figure}

However, the uniform motion is expected to break down even for the
defect-free wire when the external field or current density is large.
For the vortex wall, when a current/field is
turned on, the wall moves along the wire, and at the same time
the vortex core also tends to move perpendicular to the wire. A
restoring force \cite{Miltat3} may stop this perpendicular movement and
keep the vortex core inside the wire only when the current/field is
sufficiently small. In this case, the wall motion remains steady
and the velocity is given by Eq.~(2)
as shown in Fig.~(1). The dynamics of the vortex wall at a higher
current/field will be discussed in next section. For the transverse
wall, if one applies a larger current/field, the uniform motion becomes
unstable. In fact, the wall will constantly deform during its
motion and the velocity is highly non-uniform, known as Walker's
breakdown \cite{Walker}. In the film we consider here, we choose the applied
field and current far below the critical values for
Walker's breakdown. We do not intend to describe the complicated
dynamics after the breakdown in this paper.

\subsection{Domain wall motion in the presence of defects}

While there are many types of defects to pin a domain wall, we
restrict our discussion on the defect induced by the roughness on
the edges of the wire width. Specifically, we periodically remove
one grid at each edge as described above in otherwise a defect-free
wire.

Let us start with the transverse wall. When a current is applied,
the domain wall begins to move and in the meanwhile the wall width
shrinks. The wall shape remains similar to the original one. If the
current is small, the wall eventually stops. The displacement of the
center of the wall is proportional to the current density and the
magnitude could be as large as several tens of nanometers. The
terminal average velocity is zero. When the current exceeds a
critical value $J_{t_1}$, as shown in Fig.~(2a), the domain wall is
able to overcome the pinning and can sustain a significant velocity,
which is oscillating corresponding the defect positions (not shown
here). In fact, the average velocity are smaller but still close to
$-c_J/\alpha$ once the current density is larger than the critical
current. It is understood that part of the effect of the defects has
been compensated by the deformation of the domain wall and the
average velocity has not been reduced significantly.

One interesting feature of the transverse wall is that the wall
velocity as a function of the current is {\em hysteretic}. If the
wall reaches the critical current, $J_{t_1}$, the wall begins to
move with an average velocity, shown in Fig.~(2a). Then after the
wall is virtually depinned ($J\geq{J_{t_1}}$) and a uniform wall
velocity is established, one begins to reduce the current. It is found
that the domain wall remains to have a finite velocity even if the current
density has reduced to a value lower than $J_{t_1}$. One needs to further
reduce the current density until reaching a second critical
current, $J_{t_2}$, in order to completely stop the domain wall, see
Fig.~(2a). The hysteretic feature may be qualitatively understood
in terms of the competition between the wall kinetic energy and
defect potential \cite{Li}. When the current density is reduced from
a higher value, the velocity of the wall remains large and it would
be difficult for defects to pin the wall. On the other hand, when we
increase the current density from a smaller value, the wall is
originally at zero velocity and thus the pinning is more effective.

\begin{figure}
\centering
\includegraphics[width=8.5cm]{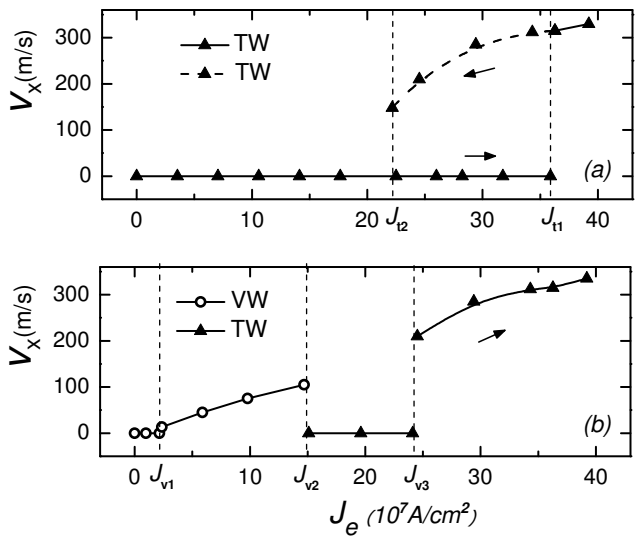}
\caption{The dependence of average domain-wall velocity on the
current and wall structures. In (a), solid line shows the velocity
of transverse wall by applying a current from the static state
($v_{x}(0)=0$,$J_{e}(0)=0$); dashed line represents the velocity by
applying a reduced current from a depinned state
($v_{x}(0) \neq0,J_{e}(0)=J_{t_{1}}$). In (b) the open circle line shows
the average velocity of vortex wall, which depins at $J_{v_1}$ and
transforms to a transverse wall at $J_{v_2}$; the up-triangle line
represents the velocity of the transverse wall after the
transformation: upon $J_{v_3}$, the transformed wall keeps the same
velocity as in (a) with corresponding currents.}
\end{figure}

The motion of the vortex wall is far more complex than that of
transverse wall, because the transformation happens when the applied
current/field is high. As in the transverse domain wall, one needs a
critical value of the current to maintain a finite terminal velocity
with vortex core inside the wall. However, the most interesting
feature is that the critical current for the vortex wall, $J_{v_1}$,
is far smaller than $J_{t_1}$, i.e., the vortex wall is less
sensitive to the defect potential. As shown in Fig.~(2b), the vortex
wall begins to move at a current density an order of magnitude
smaller than that for the transverse wall. This is somewhat
understandable because the effects of edge roughness on a vortex
wall is quite small. In the previous work \cite{Takashi}, it has been
shown that the similar defect may trap the domain wall when being
present at the center of vortex core, and it has almost no effects
to the wall if staying at the edge.

When the current continues to increase, the average velocity of the
wall proportionally increases. In the meanwhile, the center of the
vortex core moves closer to the edge. At a second critical
$J_{v_2}$, the vortex reaches the boundary and the vortex wall is
transformed into a transverse wall. Because the critical current for
the transverse wall $J_{t_1}$ is much larger, the wall motion stops.
These features seem to agree with the experimental observation
\cite{Klaui}. If the applied current is further increased upon to
the third critical value $J_{v_3}$ which is smaller than $J_{t_1}$,
the wall can still keep a finite velocity after the transformation
is completed. And this average velocity just equals to the terminal
average velocity that the transverse wall may sustain after the
current is reduced from higher level, i.e.$J_{t_1}$ in Fig.~(2a). It
is also explained as the hysteretic feature of the transverse wall
we described above: the kinetic energy gained before the wall
transformation helps the transformed wall sustain a significant
velocity.

\section{Conclusions}

We have investigated the domain wall dynamics for vortex and
transverse walls by using the spin torque model. Interestingly, while the
terminal velocity is independent of the wall structure for a steady
wall motion in defect-free wire, the depinning current of the vortex
wall is an order of magnitude smaller than the transverse wall in
the presence of defects. However, the vortex wall is not as stable
as the transverse wall during its motion. The vortex core may be
annihilated at the edges and the transformation to the transverse
wall occurs. These conclusions seem to explain the experimental
observation \cite{Klaui}. This work is supported by NSF-DMR-0314456 and the MU
research board.

\end{document}